\documentclass[Times,4pt,preprint,twocolumn,superscriptaddress]{revtex4-1}
\usepackage{amsmath,natbib,graphicx,subfigure,amssymb,graphics,amsmath,mathrsfs,CJK,color}
\usepackage{multirow,fancyhdr,color,bm,tabularx,psfrag,geometry,dcolumn,datetime}
\usepackage[colorlinks=true,linkcolor=blue,urlcolor=blue,citecolor=blue]{hyperref}
\usepackage[left]{lineno}
\def\footnoterule{\kern -1mm \hrule width 6.6cm \kern 2.2mm}%
\usepackage{tikz,xcolor,hyperref}
\definecolor{lime}{HTML}{A6CE39}
\DeclareRobustCommand{\orcidicon}{%
    \begin{tikzpicture}
    \draw[lime, fill=lime] (0,0)
    circle [radius=0.16] node[white]
   {{\fontfamily{qag}\selectfont \tiny ID}};\draw[white, fill=white] (-0.0625,0.095)
    circle [radius=0.007];
    \end{tikzpicture}
    \hspace{-2mm}}
\foreach \x in {A, ..., Z}
{\expandafter\xdef\csname orcid\x\endcsname{\noexpand\href{https://orcid.org/\csname orcidauthor\x\endcsname}{\noexpand\orcidicon}}}

\begin{document}

\title{\LARGE Left-handedness in the balanced/unbalanced resonance conditions of a quantized composite right-left handed transmission line}

\author{Xiao-Jing Wei }
\affiliation{Department of Physics, Faculty of Science, Kunming University of Science and Technology, Kunming, 650500, PR China}

\author{Shun-Cai Zhao\textsuperscript{\orcidA{}}}
\email[Corresponding author: ]{ zhaosc@kmust.edu.cn }
\affiliation{Department of Physics, Faculty of Science, Kunming University of Science and Technology, Kunming, 650500, PR China}


\begin{abstract}
Left-handedness signifies negative permittivity (\(\varepsilon_r\)) and permeability (\(\mu_r\)) in the same frequency band. The \(\varepsilon_r\) and \(\mu_r\) are evaluated in a quantized composite right-left handed transmission line (CRLH-TL), and the frequency band for left-handedness is also valuated in the balanced resonance (\( L_r C_l\!=\!L_l C_r \)) and unbalanced resonance(\(L_l C_r \!\neq\! L_r C_l\)) cases in the displaced squeezed Fock state. The results show that the balanced resonance plays an important role in bandwidth and achieving for left-handedness. These displays some quantum mechanical behaviors and proposes a new potential approach to wider frequency band left-handedness for the quantized CRLH-TL.
\begin{description}
\item[PACS: ]{42.50.Gy, 78.20.Ci, 81.05.Xj}
\item[Keywords:]{Left-handedness; wider frequency band; the balanced resonances condition; quantized CRLH-TL}
\end{description}
\end{abstract}

\maketitle
\section{Introduction}

During the infancy of metamaterials when emerged from the first experimental demonstration for a left-handed (LH) structure\cite{1Shelby2001Experimental}, the vast majority of groups were involved in the research of the fundamental properties of LH media predicted by Veselago in 1968\cite{2Veselago1968The}. And some interesting properties\cite{3Feise2002Effects,4Zhao2009LEFT,5Aydin2007Subwavelength,Du2006Quantum,Leonhardt2007}, such as reversals of both Snell's law and Doppler shift\cite{2Veselago1968The}, amplification of evanescent waves\cite{3Feise2002Effects}, subwavelength focusing\cite{5Aydin2007Subwavelength}, etc, are revealed by the technologies of artificial composite metamaterials\cite{1Shelby2001Experimental,62003Natur.423...22P}, photonic crystal structures\cite{7Cubukcu2003Electromagnetic}, photonic resonant materials\cite{8Thommen2006Electromagnetically,9Zhao2017The} and transmission line (TL) simulation\cite{10Caloz2002Application,11Caloz2006Dual,12doi:10.1002/pssb.200674510,ModeWang,2019Messinger}. But the artificial composite metamaterials seem to be of little practical interest for engineering applications because of their exhibiting high loss and narrow bandwidth consequently. Due to the weaknesses, the TL approach\cite{10Caloz2002Application,11Caloz2006Dual,12doi:10.1002/pssb.200674510} was introduced in June 2002. The transmission line simulation structures are constituted of series capacitors $C_l$ and shunt inductors $L_l$ which intend to provide left-handedness due to the simultaneously negative permittivity (\(\varepsilon_r\)) and permeability (\(\mu_r\)). Materials, as a consequence of these double negative parameters, are characterized by antiparallel phase and group velocities, and negative refractive index. However, as an electromagnetic wave propagates along these structures, the associated currents and voltages induce other natural effects: the magnetic fluxes are induced and therefore a series inductance $L_r$ is present when the currents flow along $C_l$, in addition, the voltage gradient existing between the upper conductors and the ground plane is equivalent to a shunt capacitance $C_r$. As a consequence, a purely LH structure does not exist even in a wider frequency band owing to the proliferation of ($L_r$, $C_r$) which results in the characteristic different from those generated by the series capacitors $C_l$ and shunt inductors $L_l$. Since a real LH structure necessarily includes ($L_r$, $C_r$) contributions in addition to the ($L_l$, $C_l$) reactance\cite{13Sanada2004Characteristics}, the concept of CRLH-TL based TL was introduced by Caloz\cite{14Caloz2003Invited} and the left-handedness only existed in the microwave frequency band.

Although CRLH-TLs have provided some novel structures and devices, such as couplers\cite{15Nguyen2007Generalized}, band pass filters\cite{16De2012Investigation}, antennas\cite{17Segoviavargas2013Quad}, and multi-frequency devices\cite{18Shamaileh2011NON} in this limited frequency band, in terms of the potential application, it is still a great challenge so far to achieve left-handedness in a much more wider frequency band. Not only that, but the CRLH-TL's microscale approaches the Fermi wavelength, the quantum effects on the CRLH-TL\cite{19Zhao2017Negative,20Zhao2017Quantum,21Wei2017The,22QiXuanWu,andp201900495} should be taken into account similarly to the mesoscopic circuits\cite{23Flores2001Mesoscopic,24Zhang2004Quantum,25Ji2006The,26Ji2008Dynamical}. Some novel features\cite{19Zhao2017Negative,20Zhao2017Quantum,21Wei2017The,22QiXuanWu} have been revealed in the quantized CRLH-TL as compared to the macro CRLH-TL.

With achieving a wider frequency band left-handedness in mind, we discussed the feasibility of simultaneous negative \(\varepsilon_r\), \(\mu_r\) within the same frequency band in the displaced squeezed Fock state, and compared the ranges of frequency bands in the balanced resonance case (\( L_r C_l\!=\!L_l C_r \))\cite{27Caloz2004A,28Sanada2004Characteristics} with those under the unbalanced resonances condition (\(L_r C_l\! \neq\!L_l C_r\)). It found that a wider frequency band left-handedness was achieved in the balanced resonance condition. Therefore, a new approach to left-handedness in a wider frequency band may be provided in this scheme.

And this paper is organized as follows. Firstly, the travelling current was quantized in the unit-cell circuit of the CRLH-TL, and the permittivity $\varepsilon_r$ and permeability $\mu_r$ were deduced in the displaced squeezed Fock state. Then we discussed the left-handedness in the displaced squeezed Fock state. And the summary and conclusions were presented in the end of this paper.

\section{Effective permittivity $\varepsilon_r$ and permeability $\mu_r$ in the quantized CRLH-TL}

\begin{figure}[!t]
\centerline{\includegraphics[width=0.75\columnwidth]{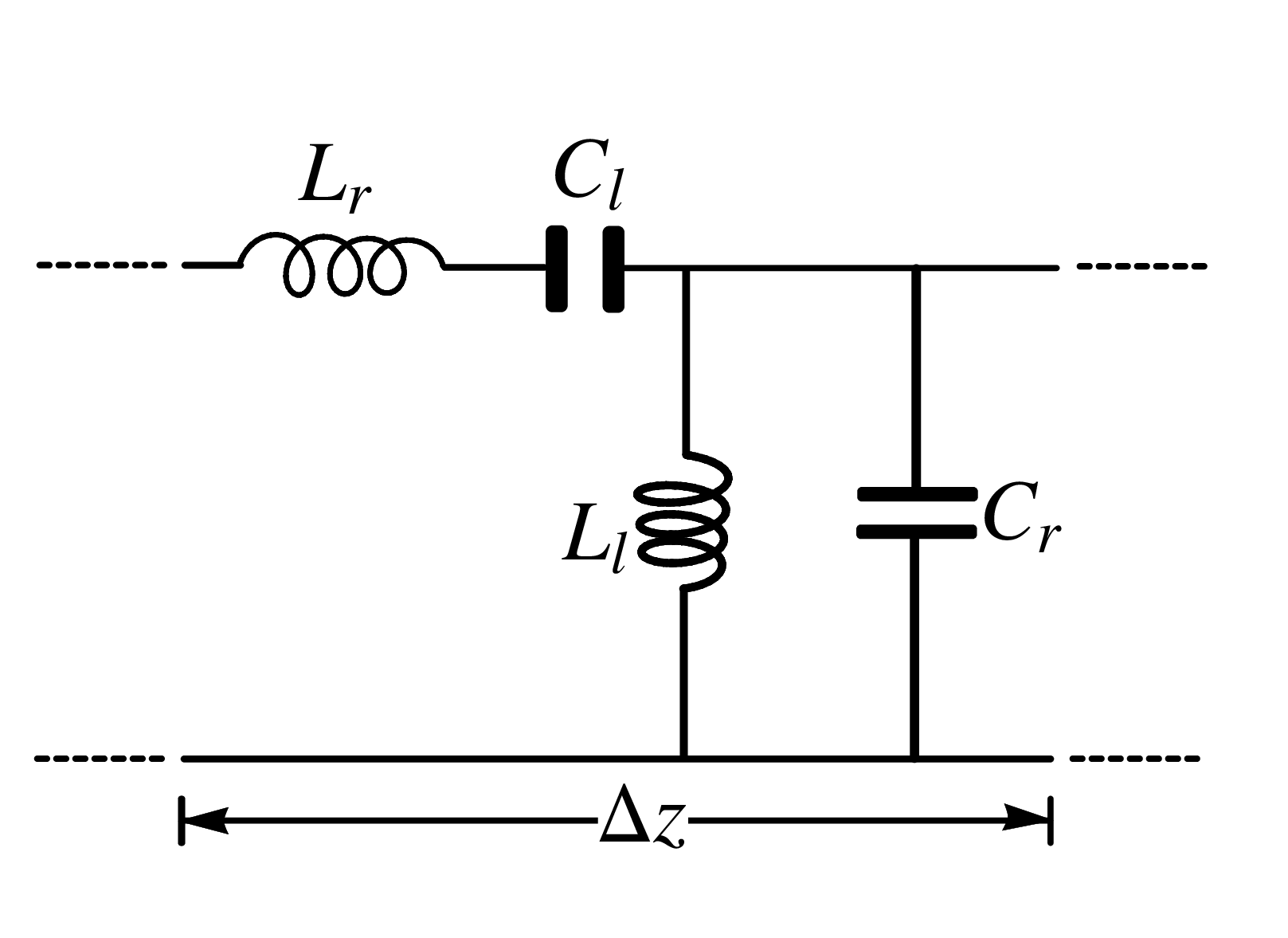}}
\caption{ Schematic diagram of equivalent unit-cell circuit of the quantized CRLH-TL.}
\label{fig1}
\end{figure}

Fig.1 shows the equivalent unit-cell circuit model\cite{29Caloz2004Transmission} for this proposed CRLH-TL, which consists of series capacitor $ C_l$ and inductance $ L_r$, shunt inductance $L_l$ and capacitor $C_r$\cite{30Eleftheriades2003Transmission}. Generally, the series capacitor $ C_l$ and inductance $L_l$ are called left-handed parameters, $C_r$ and $ L_r$ are the  right-handed parameters. And the dimension $ \triangle z$ of the equivalent unit-cell circuit is much less than the wavelength at operating frequency. The permittivity $ \varepsilon_r $, permeability $\mu_r $ are read as  \(\varepsilon_r\) = \({C_r} - \frac{1}{ \omega ^2 L_l}\), \(\mu_r\) = \({L_r} -  \frac{1}{ \omega ^2 C_l}\)\cite{31Lai2004Composite} for an unit-cell circuit in Fig.1, respectively. Hence, the Kirchhoffs' voltage and current laws for time t propagating along z direction for this unit-cell circuit in Fig.1 can be written as follows,

\begin{align}
u(z,t)&=i(z,t)[\frac{1}{j\omega\frac{C_l}{\Delta{z}}}+j\omega{L_r}\Delta{z}]+u(z+\Delta{z,t})
\end{align}

\begin{align}
i(z,t)&=i(z+\Delta{z,t})+ [ \frac{1}{ j\omega \frac{L_l}{\Delta{z}} }+ j\omega\frac{C_r}{\Delta{z}}]u(z+\Delta{z,t}).
\end{align}
\noindent in which $ u(z, t) $ stands for the voltage, $i(z, t)$ is the current, and $ \omega $ is the angle frequency. When $\Delta z \rightarrow 0$, the above equations together with the additional equations $ i(z, t) = C \frac{\partial u(z,t)} {\partial t} $ and $ C = C_r + C_l$ bring out the forward plane-wave solutions,

\begin{align}
u(z,t)&=A e^{-j( \omega t-\beta z )}+A^*  e^{j(\omega t-\beta z)},\\
i(z,t)&=j\omega C [ A^* e^{j(\omega t-\beta z)}-A e^{-j(\omega t-\beta z) }].
\end{align}

\noindent with $ \beta \!=\! \sqrt{(\frac{\omega}{\omega_r})^2 + (\frac{\omega _l}{\omega})^2 -k {\omega_l}^2 } $, where $\omega _r \!=\!\frac{1}{\sqrt{L_r C_r}} $, $\omega_l \!=\!\frac{1}{\sqrt{L_l C_l}}$, $k\!=\!L_r C_l + L_l C_r $. In Eq.(3)and Eq.(4), $A^*$ is the complex conjugate of A for the normalization purpose. In order to simplify Eq.(3)and Eq.(4), we introduce two functions $\xi (z, t)$ and $ \eta (z,t) $ as follows,

\begin{align}
u(z,t)&=\xi (z,t),\\
i(z,t)&=\frac{\omega ^2 }{z_0}  \eta (z,t).
\end{align}

\noindent  where $z_0$ stands for the length of per unit circuit, and we assume that $\omega$, the unit length of circuit $z_0$ take the fixed values. Then, $\eta (z, t)$ and $\xi(z, t) $ differential $t$ are
\vskip -4mm
\begin{align}
\frac{\partial \xi (z,t)}{\partial t}=&\frac{\omega ^2}{C z_0} \eta(z,t),\\
\frac{\partial \eta (z,t)}{\partial t}=&-C z_0 \eta(z,t).
\end{align}

And the following relation is deduced from Eq.(7) and Eq.(8) straightforwardly,

\begin{align}
\frac{\partial}{\partial \xi(z,t)}(\frac{\partial \xi (z,t)}{\partial t})+\frac{\partial}{\partial \eta (z,t)}(\frac{\partial \eta (z,t)}{\partial t})=0
\end{align}

\noindent So, it concludes that $\eta (z, t)$ and $ \xi (z, t)$ are the canonically conjugate variables and obey the hamiltonian canonical equations as follows,

\begin{align}
\frac{\partial \xi (z,t)}{\partial t} =& \frac{\partial H}{\partial \eta (z,t)},  \\
\frac{\partial \eta (z,t)}{\partial t} =& - \frac{\partial H}{\partial \xi (z,t)}.
\end{align}

The integrals of Eq.(10) and Eq.(11) can deduce the Hamiltonian for the electro-magnetic wave with the ignored integral initial constant as follows,

\begin{align}
\textit{H}=\frac{\omega ^2}{2C z_0}\eta^2(z,t)  +  \frac{1}{2} C z_0 \xi^2 (z,t),
\end{align}

If we incorporate Hermitian operators of $  \hat{ \eta }(z, t) $ and $ \hat{\xi} (z, t) $ and require them to satisfy the following commutation relation,

\begin{align*}
[\hat{\eta},\hat{\nu}]=&j \frac{C z_0}{\omega}[A e^{-j(\omega t-\beta z)}+ A^* e^{j(\omega t-\beta z)},\nonumber\\
                       &A^* e^{j(\omega t-\beta z)}-A e^{-j(\omega t-\beta z)}]\\
                      =&j\frac{2 C z_0}{\omega}[\hat{A},\hat{A^*}]=j\hbar,
\end{align*}

\noindent Then, the quantized unit-cell circuit can be obtained via the definitions,

\begin{align*}
\hat{A}=&\hat{a}\sqrt{\frac{\hbar \omega}{2C z_0}} ,   \\
\hat{A^*}=&\hat{a^*}\sqrt{\frac{\hbar \omega }{2C z_0 }}.
\end{align*}

\noindent Where the above two operators $  \hat{ \eta }(z, t) $  and $ \hat{\xi} (z, t) $ are defined as follows,

\begin{align}
\hat{\xi}(z,t)=&\sqrt{\frac{\hbar \omega }{2 C z_0}}[\hat{a}e^{-j(\omega t-\beta z)}+{\hat{a}}^{\dag}  e^{j(\omega t-\beta z)}] ,  \\
\hat{\eta}(z,t)=&j \sqrt{\frac{\hbar \omega }{2 C z_0}}[{\hat{a}}^{\dag}  e^{j(\omega t-\beta z)}+\hat{a}e^{-j(\omega t-\beta z)}].
\end{align}

Then the quantum Hamiltonian of the unit-cell circuit can be written as \( \hat{H} = \hbar \omega ({\hat{a}}^{\dag} \hat{a} + \frac{1}{2}) \) which is analogous to the oscillator's Hamiltonian operator in Schr$\ddot{o}$dinger representation. As can be seen from Equations (13) and (14), the quantum Hamiltonian operator \( \hat{H}\) contains operators \(\hat{\xi}(z,t)\) and \(\hat{\eta}(z,t)\), which both contain the parameter \(\beta\). The ingredients of the parameter \(\beta\) has parameter k, which has the expression $k\!=\!L_r C_l + L_l C_r $. Hence, the capacitance and inductance information of the entire circuit can be found in the spectrum of the Hamiltonian \( \hat{H}\). The tilde space has been introduced to be the accompaniment of the Hilbert space\cite{321988AdPhy..37..531U}, and these two spaces form a direct product space. For every operator and state in the Hilbert space, their tilde operators and states can be defined correspondingly in the tilde space. Thus, there are operators 1 and 2 corresponding to the operators \(\tilde{1}\) and \(\tilde{2}\) in the tilde space, respectively. And they satisfy the commutation relations\cite{33JiYing2003Preparation}, $[\tilde {\hat{a}},\tilde {\hat{a}}^\dag]=1$, $[\tilde {\hat{a}}, \hat{a}]=[\tilde {\hat{a}},{\hat{a}}^\dag]=[ {\hat{a}},\tilde {\hat{a}}^\dag]=0$.
The squeezing coherent state was defined by the squeezing the vacuum, squeezing coherent states as follows\cite{33JiYing2003Preparation},

\begin{align}
|z,\xi,n\rangle=\hat{D}(z)\hat{S}(\xi)|n\rangle
\end{align}

Utilizing the displacement operator \(\hat{D}(z)\)=\(\exp(z\hat{a}^{\dag}-z^{*}\hat{a})\) and squeezing operator \(\hat{S}(\xi)\)=\(\exp(\frac{1}{2}\xi \hat{a^\dag}^{2}\)-\(\frac{1}{2}\xi^{*}\hat{a^{2}})\), we can introduce their corresponding tilde operators with the displacement parameter \(z\)=\(|z|e^{i\theta}(|z|>0,0\leq\theta<2\pi)\) and squeezing parameter $\xi=|\xi|e^{i\phi}(|\xi|>0,0\leq\phi<2\pi)$ , where \(\phi\) is the squeezing angle in the tilde space. In terms of the above definitions of displacement operator $ \hat{D}(z)$ and tilde displacement operator $\tilde{\hat{D}}(z)$, squeezing operator $\hat{S}(z)$ and tilde-squeezing operator $\tilde{\hat{S}}(z)$, and using the formula \(e^{\lambda \hat{A}}\hat{B}e^{-\lambda \hat{A}}=\hat{B}+\lambda[\hat{A},\hat{B}]+\frac{\lambda^{2}}{2}[\hat{A},[\hat{A},\hat{B}]]+\cdots\),  we can prove the following equalities\cite{34Oz1991Thermal,35Fearn1988Representations}

\begin{align*}
\hat{D^{\dag}}(z)\hat{a}\hat{D}(z)=&\hat{a}+z,   \\
\hat{D^{\dag}}(z)\hat{a^{\dag}}\hat{D}(z)=&\hat{a^{\dag}}+z^{*}, \\
\hat{S^{\dag}}(\xi)\hat{a}\hat{S}(\xi)=&\hat{a}\cosh|\xi|+\hat{a^{\dag}}e^{i\phi}\sinh|\xi|, \\
\hat{S^{\dag}}(\xi)\hat{a^{\dag}}\hat{S}(\xi)=&\hat{a^{\dag}}\cosh|\xi|+\hat{a}e^{-i\phi}\sinh|\xi|.
\end{align*}

When the mesoscopic CRLH-TLs operate in the displaced squeezed Fock state in Heisenberg picture, we can deduce the equations for $\varepsilon_r $, and $\mu_r$\cite{31Lai2004Composite} via the current quantum fluctuation $ \langle (\Delta i)^2 \rangle=(\langle \hat{i}^2 \rangle - {\langle \hat{i} \rangle}^2) $ through the unit-cell circuit as follows,

\begin{multline}
\resizebox{.8\hsize}{!}{$\varepsilon_r\!=\!\frac{C_l \{-4c z_0 csc(\phi) csch(2\xi){\langle (\Delta i)^2 \rangle}+[2cot(\phi)coth(\xi)+ }
{16z^2{\omega}^2{\hbar}^2(2n+1)^2({\omega}^2C_lL_r -1)}$}\\
\resizebox{.8\hsize}{!}{$\frac{+csc(\phi)+(2n+1)csc(\phi)coth^2(\xi)]tanh(\xi)\omega\hbar \}^2}{16z^2{\omega}^2{\hbar}^2(2n+1)^2({\omega}^2C_lL_r -1)}$},
\end{multline}

\begin{multline}
\resizebox{.8\hsize}{!}{$
\mu_r\!=\!\frac{L_l \{-2c{z_0}^3 csc(\phi)csch(\xi)sech(\xi){\langle (\Delta i)^2 \rangle}+}{16z^2{\omega}^{10}{\hbar}^2(2n+1)^2({\omega}^2C_rL_l-1)}
$}
\\
\resizebox{.8\hsize}{!}
{$\frac{+(2n+1){[(csc(\phi)-2cot(\phi))coth(\xi)+csc(\phi)]\omega}^5\hbar\}^2}{16z^2{\omega}^{10}{\hbar}^2(2n+1)^2({\omega}^2C_rL_l-1)}.$}
\end{multline}

\noindent where $n$ is the number of field photons corresponding to the operator $\hat{n}$ in the Hilbert space, and c is the light speed in vacuum.

\begin{table}[hbp]
\begin{center}
\caption{Parameters for the equivalent unit-cell circuit of the quantized CRLH-TL.}
\vskip 0.25cm\setlength{\tabcolsep}{2pt}
\begin{tabular}{c|c|c|c|c|c|c}
\hline
         & \(C_l (pF)\) & \(L_l (nH) \)  & \(C_r (pF)\)  & \(L_r (nH)\)  & \(\Phi \)  & \(\xi (\times10^{-3} ) \)   \\
\hline
Fig.2(a)   &72.5        & 175.5          & 12.5        & 97.3         &       -             &   \(5\)               \\
Fig.2(b)   &100.0       & 150.0          & 20.0        & 30.0         &      -              &   \(5\)               \\
Fig.3(a)   &3.5         &8.0             &2.2          &1.5           & \( \frac{\pi}{6}\)  &     -                 \\
Fig.3(b)   &4.0         & 2.0            &6.0          &3.0           & \( \frac{\pi}{6}\)  &     -                 \\
\hline
\end{tabular}
\end{center}
\end{table}

\section{Effective permittivity $\varepsilon_r$ and permeability $\mu_r$ in the balanced and unbalanced resonance conditions}

Although some interesting electromagnetic properties\cite{27Caloz2004A,28Sanada2004Characteristics} have been demonstrated in the macro CRLH-TLs under the balanced and unbalanced resonance conditions, their quantum behaviors still deserve attentions in the quantized CRLH-TLs under these two different conditions. In the following, we will implement the discussion about the effective permittivity \(\varepsilon_r\) and permeability \(\mu_r\) via the formulas (16) and (17). And the left-handedness, i.e., the simultaneous negative \(\varepsilon_r\) and \(\mu_r\) are the goals that we want to achieve in this quantized CRLH-TL. The parameters utilized in this quantized CRLH-TL are listed in Table 1, and their orders of magnitudes are referenced to Ref.\cite{36Egger2013Multimode}. And in the following simulation, one unit length of the mesoscopic CRLH-TLs will be considered and the current quantum fluctuation \(\langle (\triangle i)^2 \rangle \)=\(2.5\), and the field photon number \(n\!=\!5\) was selected in Fig.2.

\begin{figure}[!t]
\centerline{\includegraphics[width=0.45\columnwidth]{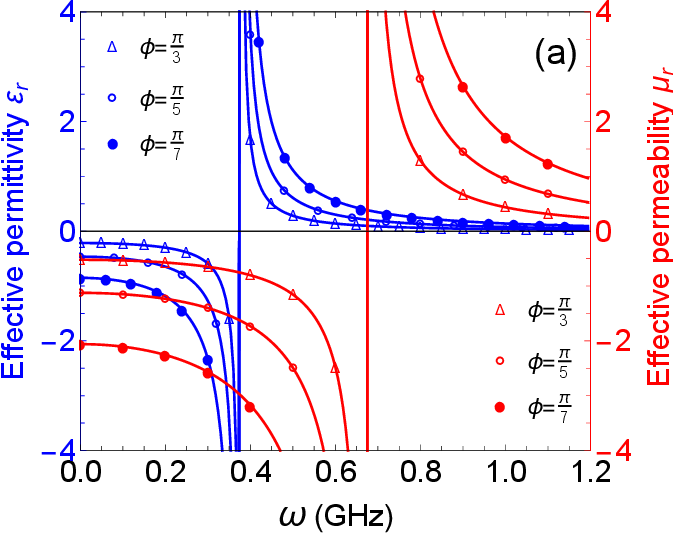}~~~~~\includegraphics[width=0.45\columnwidth]{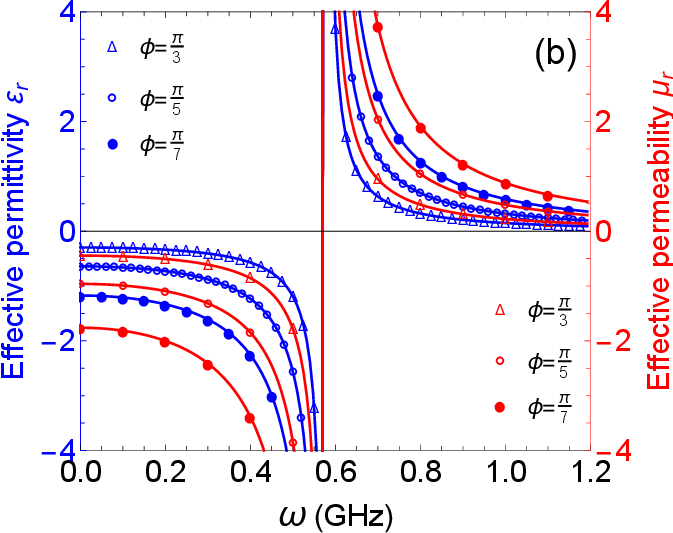}}
\caption{ (Color online)The effective permittivity \(\varepsilon_r\) and permeability \(\mu_r\) versus the frequencies \(\omega\) implemented by different squeezing angles \(\phi\) in the unbalanced (a) and balanced (b) resonance cases, respectively. }
\label{fig2}
\end{figure}

As mentioned before, the simultaneous negative $\varepsilon_r$ and $\mu_r$ in the same frequency band means the left-handedness. In Fig.2, we plot the effective permittivity $\varepsilon_r$ (curves with blue symbols) and permeability $\mu_r$ (curves with red symbols) versus the frequencies \(\omega\) in the unbalanced (a) and balanced (b) cases, respectively. And the squeezing angle \(\phi\) acts the regulatory parameter. Fig.2(a) shows that the frequency ranges for negative effective permittivity $\varepsilon_r$ are smaller than those for effective permeability $\mu_r$ with the same squeezing angle \(\phi\), and the frequency range for simultaneous negative value is broadened with the increments of the squeezing angle \(\phi\) in the case of unbalanced resonance. It notes that the negative effective permittivity $\varepsilon_r$ presents about in the frequency ranges (0, 0.4 G\(Hz\)) while the negative effective permeability $\mu_r$ is  about in the frequency ranges (0, 0.6 G\(Hz\)) when the squeezing angle \(\phi\) is set by \(\frac{\pi}{7}\), \(\frac{\pi}{5}\), \(\frac{\pi}{3}\), respectively. And the negative effective permittivity $\varepsilon_r$ and permeability $\mu_r$ co-occurrence in the range (0, 0.4 G\(Hz\)). The blue curves indicate that the left-handedness is easy to implemented with a larger squeezing angle \(\phi\) in the displaced squeezed Fock state in Fig.2(a).

In Fig.2(b), the parameters for the circuit components are coincide with \( L_r C_l\!=\!L_l C_r \) ( the balanced resonances condition ) in Table 1, and other parameters take the same values to those in Fig.2(a). The distinctive characteristic of Fig.2(b) is the frequency bands of blue curves, and the frequency bands of the blue curves move right and they are almost identical to the red curves. And the negative values for ($\varepsilon_r$, $\mu_r$) decrease with the increments of squeezing angle \(\phi\) by \(\frac{\pi}{7}\), \(\frac{\pi}{5}\), \(\frac{\pi}{3}\). Comparing the frequency bands of simultaneous negative in Fig.2(a) and (b), it concludes that the balanced resonance condition is positive for left-handedness due to the wider frequency bands achieved in the balanced resonance condition.

\begin{figure}[!t]
\centerline{\includegraphics[width=0.45\columnwidth]{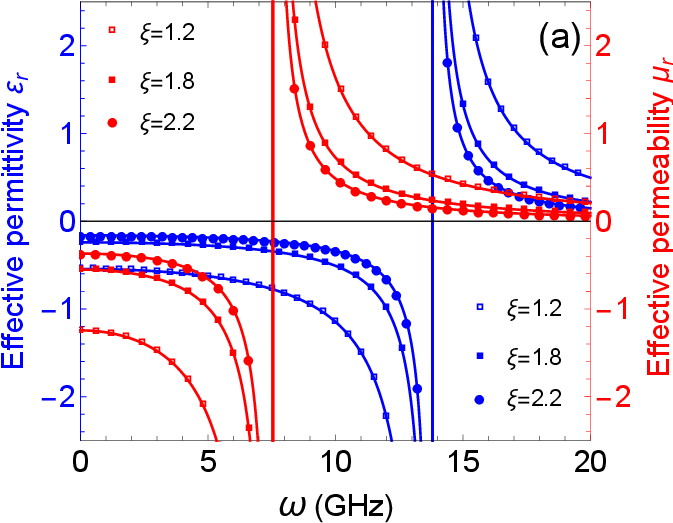}~~~~~\includegraphics[width=0.48\columnwidth]{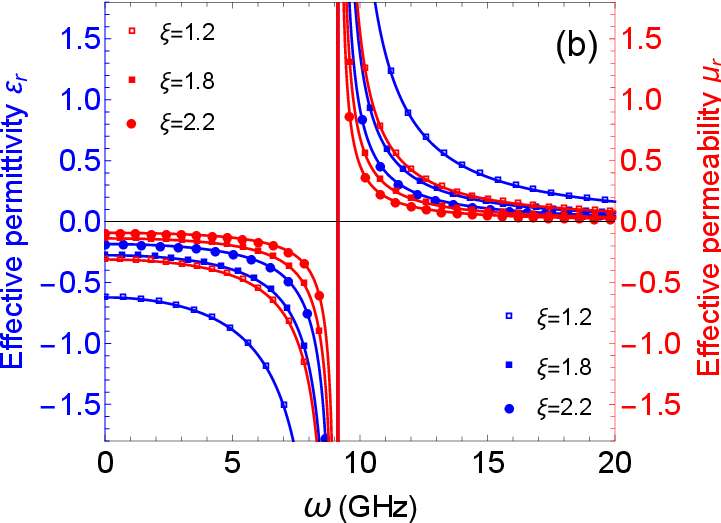}}
\caption{ (Color online) The effective permittivity \(\varepsilon_r\) and permeability \(\mu_r\) versus the frequencies \(\omega\) implemented by different squeezing parameters \(\xi\) in the unbalanced (a) and balanced (b) resonance cases, respectively. }
\label{fig3}
\end{figure}

In the displaced squeezed Fock state, there is another adjustable parameter, the squeezing parameter \(\xi\). Fig.3(a) shows the effective permittivity \(\varepsilon_r\) (curves with blue symbols) and permeability \(\mu_r\) (curves with red symbols) versus the frequencies \(\omega\) regulated by different squeezing parameters \(\xi\). And the parameters of series capacitor $C_l$ and inductance $L_r$, shunt inductance $L_l$ and capacitor $C_r$ in Table 1 constitute the inequality relation \(L_l C_r\! \neq\! L_r C_l\), other parameters are the same to those in Fig.2.
Differently, the frequency ranges of negative effective permittivity \(\varepsilon_r\) are larger than those for permeability \(\mu_r\) in Fig.3 (a). And the frequency range has an order of magnitude larger than those in Fig.2(a). The frequency ranges of negative permeability \(\mu_r\) are about [0, 7GHz], and about for [0, 13.5GHz] for effective permittivity \(\varepsilon_r\) in Fig.3(a).
Therefore, left-handedness is achieved about in the frequency range [0, 7GHz] in the unbalanced resonance case. The absolute values of both negative effective permeability \(\mu_r\) and permittivity \(\varepsilon_r\) decrease with the increments of the squeezing parameter \(\xi\) from 1.2, 1.8 to 2.2, which indicates a prominent squeezing effect on the left-handedness.

However, in the balanced resonance case \(L_l C_r \!=\!L_r C_l\) (listed in Table 1) in Fig.3(b), the frequency ranges for negative permeability \(\mu_r\) and permittivity \(\varepsilon_r\) both move to the location \(\omega\)\(\approx\)10GHz, while their amplitudes are almost stationary. And the frequency ranges for simultaneous negative (\(\varepsilon_r\), \(\mu_r\)) in Fig.3(b) are larger than those in Fig.3(a). The results achieved in Fig.2(b) and Fig.3(b) come to the same conclusion that the balanced resonance condition may be a good choice for the wider frequency band left-handedness except the facility of experimentally selecting the balanced resonance circuit elements.

Before concluding above work, some items should be emphasized here. Firstly, the simultaneous negative effective permittivity \(\varepsilon_r\) and permeability \(\mu_r\) i.e., the left-handedness can be achieved and adjustable in the quantized CRLH-TL. And wider frequency band left-handedness implemented in the balanced resonance condition (\( L_r C_l\!=\!L_l C_r \)) is the typical feature of this scheme. Secondly, in experimental research, it should be superconducting\cite{37Salehi2005Analysis} for CRLH-TL to reach the quantum regime. And the Josephson Junction Arrays can generate the left-handed transmission line\cite{36Egger2013Multimode} due to their compact way to engineer sizable inductances\cite{38Pop2014Measurement,392011PhRvB..83a4511H}. Therefore, the Josephson Junction Arrays may be a potential candidate for the quantized CRLH-TL in the coming experiments.

\section{Conclusion}

In the present work, we focused on the simultaneous negative effective permittivity \(\varepsilon_r\) and permeability \(\mu_r\), i.e., left-handedness of the quantized CRLH-TL in the displaced squeezed Fock state under the balanced and unbalanced resonance conditions. The results show that the simultaneous negative (\(\varepsilon_r\), \(\mu_r\)) can be achieved by different squeezing angles \(\phi\) and displacement parameters \(\xi\), and the frequency ranges regulated by the displacement parameters \(\xi\) have one order of magnitude larger than those controlled by the squeezing angles \(\phi\). What's more, the frequency ranges for simultaneous negative ( \(\varepsilon_r\), \(\mu_r\)), i.e., the left-handedness in the balanced resonances (i.e., \( L_r C_l=L_l C_r \)) are much wider than those in the unbalanced resonances(i.e., \(L_l C_r \neq L_r C_l\)). The results demonstrate the quantum-mechanical behaviors of the left-handedness in the quantized CRLH-TL, and provide a possible approach to the left-handedness with wider frequency bands. The wider frequency band left-handedness may attract some novel research interests in the potential applications.

\section*{Conflict of Interest}

The authors declare that they have no conflict of interest. This article does not contain any studies with human participants or animals performed by any of the authors. Informed consent was obtained from all individual participants included in the study.

\section*{Acknowledgments}

We thank the financial supports from the National Natural Science Foundation of China ( Grant Nos. 61565008 and 61205205 ), and
the General Program of Yunnan Applied Basic Research Project, China ( Grant No. 2016FB009 ).

\section*{Author contribution statement}

Zhao S. C. conceived the idea and guided the analysis and revised the paper. Wei X. J. performed the numerical computations and wrote the draft.

\bibliography{reference}
\bibliographystyle{unsrt}
\end{document}